\documentclass{article}

\usepackage[english]{babel}

\usepackage[letterpaper,top=2cm,bottom=2cm,left=3cm,right=3cm,marginparwidth=1.75cm]{geometry}

\usepackage{amsmath}
\usepackage{graphicx}
\usepackage{cite}

\usepackage[colorlinks=true, allcolors=blue]{hyperref}

\title{\bf Theory Motivation:\\
What measurements are needed?}
\author{Alexander Lenz}

\begin{document}
\maketitle

\begin{abstract}
I give a theory motivation for future measurements in quark flavour physics, trying to identify observables, which are less familiar, but nevertheless interesting and promising.

\end{abstract}

\tableofcontents

\section{Introduction}
The organisers of the 2021 edition of {\it Heavy Quarks and Leptons (HQL)} at Warwick
University \footnote{See the INDICO page for a list of excellent talks and of mine: 
\href{https://indico.cern.ch/event/1073157/}{https://indico.cern.ch/event/1073157/}}
were asking me to give a theory motivation of measurements that should be pursued in
future. The {\it HQL} conference series is covering a very broad field and due to my
ignorance I will only concentrate on observables related to weak decays of hadrons
containing a $b$ or a $c$ quark. However, even the remaining field is too large to be
covered in a 25 minutes talk, thus I will try to identify some observables, which are less
in the bright spotlight of  our community, but that might nevertheless provide some
interesting insights. 
{\color{blue} 
Whenever I am possibly  adding something  new to the current discussion
of future observables I will use blue colour for the text.}
In addition I am classifying the discussed observables into three categories:
\begin{itemize}
    \item[(A)] The Usual Suspects
    \item[(B)] Not so well known 
    \item[(C)] Surprises/Underdogs
\end{itemize}
\section{The Usual Suspects}
\label{sec:usual}
The observables falling in this category are described in numerous reports like
in the Report {\it Opportunities in Flavour Physics at the HL-LHC and HE-LHC}
\cite{Cerri:2018ypt} or the Belle II physics book \cite{Belle-II:2018jsg}
but also in many of the talks presented at HQL 2021
- I will mostly try to avoid 
these observables, except I think there is an additional point to be made.
\subsection{Flavour Anomalies}
First and foremost our field probably requires an independent experimental check, by e.g. ATLAS, CMS and Belle II 
(see the talk of Alessandro Cerris covering ATLAS and CMS)
of  the flavour anomalies
found by the LHCb Collaboration. 
The consequences of these results are very far reaching and they have to be independently
confirmed. Next on the list there is the need for a higher experimental precision particularly in the
theoretically clean observables, like $R_K $  and friends. Moreover it would be important  to extend 
these kind of measurements to different decay channels, e.g. baryonic ones like 
$\Lambda_b \to \Lambda^{(*)} \ell^+ \ell^-$ \cite{Bordone:2021usz},
but also rare ones like $B \to K \nu \bar{\nu}$ \cite{Browder:2021hbl,Belle-II:2021rof}, see e.g. the talks of Abdul Basith, Moritz Bauer, Marzia Bordone, Mark Smith.
\subsection{Determination of CKM elements}
Our bread and butter activity of determining precisely  the CKM matrix might, however,  also hold
some  surprises.
\begin{itemize}
\item[$\diamond$] There is still the ongoing puzzle that the inclusive and exclusive
determination of $V_{cb}$ and $V_{ub}$,
  do not agree, see the talk of Paolo
     Gambino.
\item[$\diamond$] The Cabibbo anomaly corresponds to the fact that  the sum
     $ |V_{ud}|^2 +  |V_{us}|^2 + |V_{ub}|^2 $  differs from one, see e.g.
     \cite{Grossman:2019bzp} for the first mentioning of the expression {\it Cabibbo anomaly} in
     a paper title and the talk of Martin Gorbahn.
\item[$\diamond$] A direct determination of the CKM elements including the top quark, $V_{tx}$,
     without relying on the unitarity of the CKM matrix, see also the talk of Gagan Mohanty.
\item[$\diamond$] CP violation in the $B_d$ system  - denoted by $\sin 2 \beta$ - was
     discovered as a large effect in the decay $B_d \to  J / \psi K_S$ \cite{BaBar:2001pki,Belle:2001zzw}.
     Later on the measurement of  $\sin 2 \beta_s$  in $B_s \to J / \psi \phi$ was considered to
     be a promising candidate for a Null test of the SM. Due to some huge experimental progress
     (see the talks by Alessandro Cerri and Alex Gilman) the HFLAV \cite{HFLAV:2019otj} value of $\phi_s = - 2 \beta_s$
     now reads
     \begin{equation}
         \phi_s = (- 0.050 \pm 0.019)  \, {\rm rad} = (-2.9 \pm 1.1)^\circ \, .
     \end{equation}
    {\color{blue}  Neglecting penguin contributions in the  $B_s \to J / \psi \phi$ the phase  $\phi_s $ is just given by a combination of CKM elements, see e.g.
    \cite{Artuso:2015swg}
    for which the CKM fitting groups \cite{Charles:2004jd,UTfit:2006vpt} obtain      
    \begin{equation}
         \phi_s = 
         \left\{
          \begin{array}{rcrl}
        ( - 0.03696^{+0.00072}_{-0.00084})  \, {\rm rad} & = &  (-2.12^{+0.04}_{-0.05})^\circ & \rm CKMfitter
                  \\[2mm]
        (-  0.03700  \pm 0.00104 )  \, {\rm rad} & = &  (-2.12 \pm 0.06)^\circ  & \rm UTfit
          \end{array}     
         \right. \ .
         \label{eq:betas_SM}
     \end{equation}
Contributions stemming from penguin correction are estimated to modify the values in Eq.(\ref{eq:betas_SM}) at most
by about  $1^\circ$, see e.g. \cite{Artuso:2015swg} and references therein, which is more or less identical to the
current experimental precision. In order to make use of an increasing experimental precision in $\phi_s$ further
theoretical improvements (e.g. in the line of \cite{Frings:2015eva}) or precise determinations of penguin control
channels (see e.g. \cite{Barel:2020jvf}) are mandatory. Until these theoretical improvements will not be available more
precise measurements of $\Delta \Gamma_s$ and $\Gamma_s$ \footnote{The corresponding theory predictions are obtained within the framework of the heavy quark expansion (HQE) and can be systematically improved. For the case of $\Gamma_s$ the current theory precision is higher than the corresponding experimental one.} 
from the analysis of $B_s \to J / \psi \phi$  will be more
insightful than a more precise value of $\phi_s$, which appears in many future experimental  projections.}
  \item[$\diamond$]
  In the case of the  CKM angle $\gamma$ the theory situation is much better compared to $\beta$
  and $\beta_s$. Corrections to the relation of the measured value of  $\gamma$ to a simple combination of CKM elements
  can only be of the order of $10^{-7}$ within the SM \cite{Brod:2013sga}.
  Alex Gilman presented a new experimental average of extractions of the CKM angle $\gamma$ by the LHCb Collaboration, yielding
\begin{equation}
    \gamma = \left( 65.4^{+3.8}_{-4.2} \right)^\circ \, ,
\end{equation}
which lies close to e.g. the CKMfitter value of  $\left( 65.80^{+0.94}_{-1.29} \right)^\circ $. Here experiment will
continue to produce more precise values. 
{\color{blue}  
New CP violating effects in non-leptonic tree-level decay can, however, modify
the SM relation between  $\gamma$ and the CKM elements by several degrees. It is probably less known, that such a possibility is
currently not excluded by any bounds, see \cite{Brod:2014bfa} for the first observation of this statement and 
\cite{Lenz:2019lvd} for a recent update. We will come back several times in these proceedings to the possibility and potential hints, as well as
some potential cross-checks, of BSM effects in non-leptonic tree-level decays.
}
\end{itemize}    

\subsection{Semileptonic CP asymmetries}
A flavour specific decay $B \to f$ is defined by 
\begin{eqnarray}
\langle f |{\cal H}_{eff}  |\bar{B} \rangle 
& = & 0 =
\langle \bar{f} |{\cal H}_{eff} | {B} \rangle
\, ,
\\
\langle f|{\cal H}_{eff}  | B \rangle & =& \langle \bar{f}|{\cal H}_{eff}  | \bar{B} \rangle  \, ,
\end{eqnarray}
where the second relation states that  there is no direct CP violation in these decays. Examples 
of flavour-specific decays are semileptonic decays or the decay
 $ \bar{B}_s \to D_s^+ \pi^-$.
The flavour specific CP asymmetry is then defined as 
\begin{equation}
    a_{fs}^q = \frac{\Gamma (\bar{B}_q (t)  \to f) - \Gamma ({B}_q (t)  \to \bar{f}) }
                    {\Gamma (\bar{B}_q (t)  \to f) + \Gamma ({B}_q (t)  \to \bar{f}) }
               = \left| \frac{\Gamma_{12}^q }{M_{12}^q } \right| \sin \phi_{12}^q  \, ,
               \label{ea:def_asl}
\end{equation}
with the mixing phase $\phi_{12}^q  = \arg (- M_{12}^q  /\Gamma_{12}^q )$.
Often this asymmetry is called semileptonic CP asymmetry.
The most recent SM prediction is given in \cite{Lenz:2019lvd} based on the calculations in
\cite{Beneke:1996gn,
Beneke:1998sy,
Beneke:2003az,
Ciuchini:2003ww,
Lenz:2006hd,
FermilabLattice:2016ipl,
Kirk:2017juj,
King:2019lal,
Dowdall:2019bea,
DiLuzio:2019jyq,
Davies:2019gnp}
\begin{eqnarray}
    a_{fs}^d = (-4.73 \pm 0.42) \cdot 10^{-4}  \, ,
    && 
    a_{fs}^s =  (2.06 \pm 0.18) \cdot 10^{-5}  \, ,
    \nonumber
    \\ 
    \left| \frac{\Gamma_{12}^d }{M_{12}^d } \right| =   (4.82 \pm 0.65) \cdot 10^{-3}  \, ,
    &&
    \left| \frac{\Gamma_{12}^s}{M_{12}^s } \right| =
       (4.82 \pm 0.64) \cdot 10^{-3}  \, ,
    \nonumber
    \\     
    \phi_{12}^d  = (-98\pm 19) \, {\rm mrad} 
    = (-5.6 \pm 1.1)^\circ  \, ,
    &&
    \phi_{12}^s  = (4.3 \pm  0.8) \, {\rm mrad} 
     = (0.25 \pm 0.05)^\circ    \, .
\end{eqnarray}
So far these tiny quantities have not been measured in experiment and  HFLAV \cite{HFLAV:2019otj}  states 
only relative loose bounds of
\begin{eqnarray}
    a_{fs}^d =    (-21 \pm 17) \cdot 10^{-4}  \, ,
    && 
    a_{fs}^s =  (-60 \pm 280) \cdot 10^{-5}     \, .
\label{eq:asl_HFLAV}
\end{eqnarray}
A more precise experimental investigation of flavour-specific CP asymmetries is interesting for many reasons, e.g.
\begin{itemize}
\item[$\diamond$] They provide an interesting Null test of the SM, i.e. any significant deviation from 
     the SM predictions presented above, might indicate a clear signal of new physics.  
     {\color{blue}   Note that the SM predictions might have a larger uncertainty than quoted above, see
     \cite{Lenz:2020efu}, but nevertheless a measurement of $a_{sl}^s$ with a value of the 
     order of $ 10^{-4}$ or larger is an unambiguous signal for new physics.}
    \item[$\diamond$] The result of the Dimuon asymmetry measured by the D0 Collaboration
    \cite{D0:2010sht}, might point towards enhanced values of the semileptonic CP asymmetries and of
    the decay rate difference of neutral $B_d$ mesons, $\Delta \Gamma_d$ \cite{Borissov:2013wwa}. We will come back to the
    latter quantity in Section \ref{sec:DGd}.
    \item[$\diamond$]{\color{blue}   Recently some interesting models of baryogenesis have been proposed by the 
    late Ann Nelson and her students and collaborators \cite{Elor:2018twp}
    and later further developed in \cite{Alonso-Alvarez:2021qfd,Elahi:2021jia}. In these frameworks 
    the necessary amount of CP violation to fulfill the Sakharov criteria stems from CP violating effects in $B$ or $D$ meson decays, 
    in particular in \cite{Alonso-Alvarez:2021qfd} it is the semileptonic CP asymmetries, which give 
    rise to the existence of the matter-antimatter asymmetry. Thus a more precise measurement of
    $a_{sl}^q$ might be directly related to some very fundamental properties of our Universe. Moreover
    it is very interesting to note that these models present counter examples to the common lore that the
    amount of CP violation in the CKM matrix is orders of magnitude too small to be responsible for
    baryogenesis.}

    \item[$\diamond$]  {\color{blue} Matthew Kirk discussed in his talk the possibility of the CP violating effects within
    the classical B-anomalies. Such a scenario could e.g. originate from BSM contributions to non-leptonic  
    $b \to c \bar{c} s$ transitions, see \cite{Jager:2019bgk,Jager:2017gal}, which can severely affect rare transitions like
    $b \to s \ell \ell $ and $b \to s \gamma $, but also quantities like $\tau (B_s)$, $\Delta \Gamma_s$ and $a_{sl}^s$ and
    the decay   $B_s \to J / \psi \phi$. More precise values of  $a_{sl}^s$ will thus constrain this possibility.}
    \item[$ \diamond$] We further discuss some experimental prospects of the semileptonic CP asymmetries at the very end of these proceedings, in Section \ref{sec:NLTD}.
\end{itemize}
If there are BSM effects acting only in the dispersive part of $B_s$ mixing, i.e. in the convention of \cite{Lenz:2006hd,Lenz:2011zz}, 
\begin{equation}
    M_{12}^s =  M_{12}^{s, \rm SM} |\Delta_s| e^{i \phi_s^\Delta}    \, ,
\end{equation}
and if one neglects SM penguins as well as any other potential BSM effect,
then the above discussed phases $\beta_s$ and $\phi_{12}^s$ would be modified as
\begin{eqnarray}
    -2 \beta_s^{\rm SM}   & \to & -2 \beta_s^{\rm SM}  +     \phi_s^\Delta   \, ,
    \\
    \phi_{12}^{s,\rm SM}    & \to & \phi_{12}^{s, \rm SM}  +     \phi_s^\Delta   \, .
\end{eqnarray}
Having the small SM expectations of  $\beta_s^{\rm SM} $ and $\phi_{12}^{s,\rm SM }$ in mind one might be tempted
to neglect them in the presence of a sizable new phase $ \phi_s^\Delta $ in order to find strong constraints
on the possible values of $a_{sl}^s$ from  precise  measurements of the phase $\phi_s$.
\\
In \cite{Lenz:2011zz} potential BSM contributions to the phases  $\beta_s$ and $\phi_{12}^s$ were 
studied more carefully and one finds
\begin{eqnarray}
    -2 \beta_s^{\rm Exp} & = & -2 \beta_{s,\rm Tree} ^{\rm SM}  +      \phi_s^\Delta   
      + \beta_{s, \rm Peng}^{\rm SM}  + \beta_{s, \rm Peng}^{\rm BSM} \, ,
    \\
    \phi_{12}^{s,\rm Exp}  & = &  \phi_{12}^{s, \rm SM}   +     \phi_s^\Delta    +     \tilde{\phi}_s^\Delta  \, ,
\end{eqnarray}
with $\tilde{\phi}_s^\Delta $ stemming from new effects in the absorptive part of mixing,
$ \Gamma_{12}^s = \Gamma_{12}^{s, \rm SM}  \left| \tilde{\Delta} \right| e^{-i \phi_s^{\tilde{\Delta}}} $,
which could e.g. originate from new $b \to s \tau \tau$ transitions \cite{Bobeth:2011st} or new 
non-leptonic tree-level decays \cite{Brod:2014bfa,Bobeth:2014rda,Lenz:2019lvd}.
{\color{blue}
Taking all these possibilities into account we currently do not see any strict constraint on  $a_{sl}^s$ from e.g.
$ \beta_s^{\rm Exp} $
beyond the direct experimental bounds on the semileptonic CP asymmetries quoted by HFLAV, see Eq.(\ref{eq:asl_HFLAV}), and thus any further experimental
improvement on $a_{sl}^q$ is highly desirable.
}
\subsection{CP violation in charm}
\label{sec:CPVcharm}
CP violation was measured for the first time in 2019 \cite{LHCb:2019hro} in the charm system in the quantity $\Delta A_{CP}$,
describing direct CP violation in the decays $D^0 \to K^+ K^-$ and  $D^0 \to \pi^+ \pi^-$.
Unfortunately the corresponding theory situation is still ambiguous and it is not clear yet if this result requires BSM
effects  \cite{Chala:2019fdb,Dery:2019ysp} or not \cite{Grossman:2019xcj,Cheng:2019ggx}.
Here measurements of control channels and experimental studies of e.g.  baryonic equivalents of $\Delta A_{CP}$,  see 
e.g. the talk of Fabio Ferrari  or  the recent review \cite{Lenz:2020awd} and references therein, would be very desirable.
\\
At the beginning of this year the LHCb Collaboration measured also for the first time a finite value of the mass difference of
neutral $D$ mesons with more than five standard deviations \cite{LHCb:2021ykz}. 
{\color{blue}
Unfortunately the corresponding theory
predictions based on inclusive approaches are overshadowed by an extreme GIM cancellation - a naive application of the heavy
quark expansion (HQE) leads to predictions which are orders of magnitude below the experiment. 
In  \cite{Lenz:2020efu} it was realised that the naive HQE application corresponds to assuming for the individual mixing
contributions stemming from e.g. an internal $s \bar{s}$ quark pair and an internal  $s \bar{d}$ quark pair a relative
theory uncertainty of the order of $10^{-5}$. More realistic uncertainty estimates can be achieved with the scale setting
procedure suggested in  \cite{Lenz:2020efu} - in this way the HQE prediction covers also the experimental results.
The new scale setting procedure has no effect on quantities like $B$ or $D$ meson lifetimes and the decay difference of
neutral $B$ mesons, where no GIM cancellations are present, but it might affect the theory prediction of semileptonic CP
asymmetries and enhance the corresponding theory uncertainty, see  \cite{Lenz:2020efu}.
\\
 From the experimental side the next steps will  probably include the search for CP violation in mixing, see e.g. the talk 
 of Yu Zhang. In case of a corresponding announcement I am predicting at least one theory paper arguing for the obvious 
 need of new physics effects and one arguing for the apparent existence of large non-perturbative effects.
\\
Here clearly a deeper understanding of the SM theory is necessary\footnote{The MIAPP 2022 program 
{\it CHARMING CLUES FOR EXISTENCE} can be considered as a step in that direction, see
https://www.munich-iapp.de/charmingclues.} and one way of proceeding is to consider quantities that have a simpler structure
than exclusive non-leptonic decays and where no severe GIM
cancellations are arising. In Section \ref{sec:incl} we will
therefore reconsider the status of $D$ meson lifetimes and inclusive semileptonic $D$ meson decays, following \cite{King:2021xqp}.
}
\section{Not so well known}
\label{sec:not}

\subsection{Flavour Anomalies }
There are some classes of observables related to the flavour anomalies that are less well known.
\begin{itemize}
\item[$\diamond$] Inclusive $ \overline{B}\to {X}_s{\mathrm{\ell}}^{+}{\mathrm{\ell}}^{-} $ decays are governed by the 
    same quark level transition $b \to s \ell \ell$ as the flavour anomalies, but they have a completely different structure of
    theory uncertainties than exclusive decays, see e.g.    \cite{Huber:2020vup}.
\item[$\diamond$]   $b \to s \tau^+ \tau^-$ transitions are hardly constrained by experiment, see the talk by Thanh Dong.
    Many BSM models explaining both the $b \to s \ell^+ \ell^-$ and $b \to c \ell^- \bar{\nu}_\ell$ anomalies predict large effects in 
    $b \to s \tau^+ \tau^-$, see e.g.  \cite{Cornella:2020aoq}.
   {\color{blue} New effects in  $b \to s \tau^+ \tau^-$ will also be constrained by $\Delta \Gamma_s$ and $\tau(B_s)$ and the
   semileptonic CP asymmetries \cite{Bobeth:2011st}. }
\item[$\diamond$] {\color{blue} Angular distributions of $B \to D^* \mu \nu $  or even better  $B \to D^* \tau \nu $ will be sensitive
to some NP operators that are elusive in $q^2$ distributions, see e.g. \cite{Mandal:2020htr}.
}
\item[$\diamond$]   {\color{blue} Accurate data on branching ratios and amplitude decompositions of hadronic  $B_s$ decays such as 
$B_s \to \psi \phi$ (where $\psi$ are charmonia $J / \psi, \psi (2S)$ and above), also $B_s \to D^{(*)} \bar{D} \phi$ 
are needed for a complete LCSR analysis of nonlocal effects (charm loop etc. ) in $B_s \to \phi \ell^+ \ell^-$.}

\item[$\diamond$]   {\color{blue}  $b \to d \ell^+ \ell^-$ transitions:
CKM suppressed decays like 
$B_s \to \eta \ell^+ \ell^-$,
$B \to \rho \ell^+ \ell^-$,
$B \to \omega \ell^+ \ell^-$
are not yet observed at the B-factories or at the LHC -
$B \to \pi \ell^+  \ell^-$ \cite{LHCb:2012de, LHCb:2015hsa} 
and recently
$B_s \to K^* \ell^+ \ell^-$ \cite{LHCb:2018rym}
have been observed with sizable uncertainties by LHCb.
From a  theoretical point of view these decays cannot simply be CKM-rescaled from  $b \to s \ell^+ \ell^-$, they require the determination of 
new genuine contributions, see e.g.  \cite{Rusov:2019ixr,Khodjamirian:2017fxg,Hambrock:2015wka,Ali:2013zfa}.
Interesting new obervables would further be differential distributions and CP asymmetries, but also ratios 
like $R_\pi$, suggested in \cite{Bordone:2021olx}, see also the talk of  Marzia Bordone.
Such  studies have the potential of providing valuable, complementary insight into the origin of the deviations
observed in $ b \to s \ell^+ \ell^-$ transitions.}

\end{itemize}

\subsection{$\Delta \Gamma_d$ }
\label{sec:DGd}
{\color{blue}  The decay rate of the neutral $B_d$ meson, $\Delta \Gamma_d$, is not yet measured 
  \cite{Gershon:2010wx} and the strongest experimental direct bound was so far obtained by the ATLAS
    Collaboration \cite{ATLAS:2016mln}. 
    The previously mentioned result  of the D0 Dimuon asymmetry \cite{D0:2010sht}, might point towards an 
    enhanced value of  $\Delta \Gamma_d$ \cite{Borissov:2013wwa}, which could originate in 
    BSM effects in non-leptonic tree-level decays 
     or in BSM effects in $b \to d \tau^+ \tau^-$ transitions,
     without being in conflict with other constraints, see
    \cite{Bobeth:2014rda}.}

\subsection{$B_s$ lifetime}
\label{sec:tauBs}
The lifetime of the $B_d$  and  $B_s$ mesons are measured precisely to be (HFLAV \cite{HFLAV:2019otj})
\begin{equation}
     \tau (B_d)^{\rm Exp}  = 1.519(4) {\rm ps}^{-1} \, , 
    \hspace{1cm}
   \tau (B_s)^{\rm Exp}  = 1.516(6) {\rm ps}^{-1} \, , 
    \hspace{1cm}
     \left[ \frac{\tau (B_s) }{\tau (B_d) } \right]^{\rm Exp} = 0.998 \pm 0.005  \, .
    \label{eq:taubs_exp}
\end{equation}
The most recent SM prediction \cite{Kirk:2017juj} based on the calculations in
\cite{Beneke:2002rj,Franco:2002fc,Gabbiani:2004tp} and  the Bag parameter from \cite{Kirk:2017juj} yield
also a value for the lifetime ratio close to one with a small uncertainty
\begin{equation}
   \left[\frac{\tau (B_s) }{\tau (B_d) }  \right]^{\rm HQE}= 1.0007 \pm 0.0025  \, .
    \label{eq:taubs_SM}
\end{equation}
The theory expression was hereby obtained via
 \begin{eqnarray}
   \left[  \frac{\tau (B_s) }{\tau (B_d) }  \right]^{\rm HQE}& = & 1 + \left[ \Gamma (B_d)- \Gamma (B_s)\right]^{\rm HQE}  \tau (B_s)^{\rm Exp} \, .
\end{eqnarray}
This expression has the advantage that in the HQE expansions of the total decay rate the leading term, describing
the free $b$ quark is cancelling in the the square bracket. One finds also in all the other higher order corrections an
almost  perfect cancellation of the $SU(3)_F$ breaking effects and thus a precise prediction of the ratio close 
to one. Further improvements of the experimental precision in the $B_s$ lifetime might be motivated by the following observations:
\begin{itemize}
\item[$\diamond$]  {\color{blue}
From a naive theorist perspective there seems to be an interesting discrepancy between the 
most recent determination by the ATLAS collaboration \cite{ATLAS:2020lbz},  the CMS collaboration
\cite{CMS:2020efq} and the LHCb collaboration \cite{LHCb:2019sgv,LHCb:2019nin}. All these analyses investigate
the decay $B_s \to J \psi \phi$ to extract e.g. $\Gamma_s$,  $\Delta \Gamma_s$ and   $\phi_s$.
\begin{figure}
\centering
\includegraphics[width=0.9\textwidth]{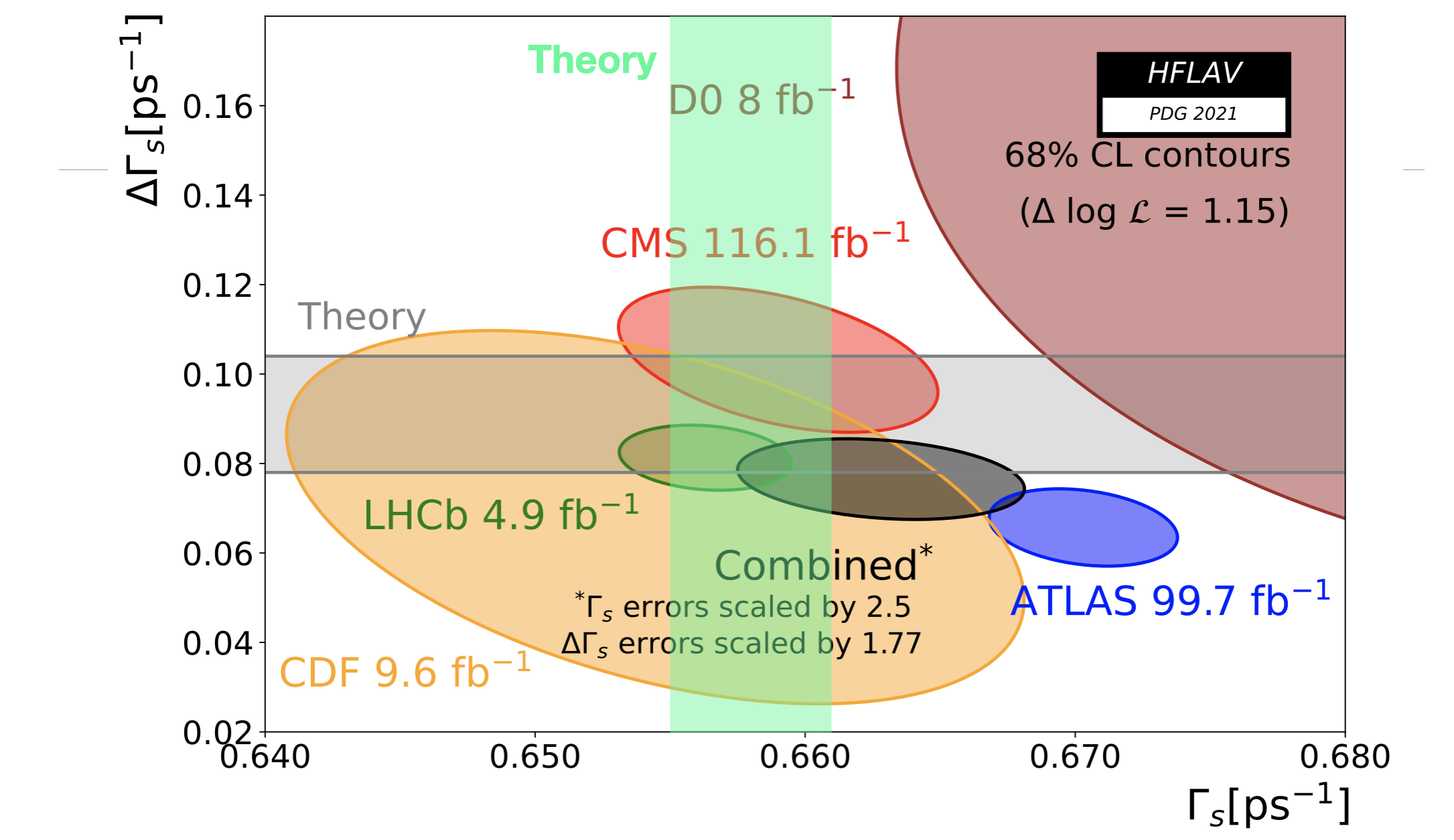}
\caption{\label{fig:bs} Experimental combination of results for $\Delta \Gamma_s$ and  $ \Gamma_s$ by HFLAV.
We have drawn by hand the SM prediction for $ \Gamma_s$  from Eq.( \ref{eq:Gs_exp}) in the colour {\it sea foam}.}
\end{figure}
The combination of their results for  $\Gamma_s$ and  $\Delta \Gamma_s$ can be found in Fig.\ref{fig:bs}, we have
added to the HFLAV plot the theory prediction for $\Gamma_s$ obtained via
 \begin{eqnarray}
 \Gamma (B_s) 
  =\frac{1 }{\tau (B_d)^{\rm Exp} } \frac{1 }{  \left[ \frac{\tau (B_s) }{\tau (B_d) }\right]^{\rm HQE}}
  = 0.658 \pm 0.003 
  \, ,
  \label{eq:Gs_exp}
\end{eqnarray}
 where we were covering the whole range of the inputs from 
 Eq.(\ref{eq:taubs_exp})
and Eq.(\ref{eq:taubs_SM})
 in order to get a conservative estimate for the range of $\Gamma_s$.}
\item[$\diamond$]   {\color{blue}
A precise experimental  and theoretical knowledge on $\tau (B_s)  / \tau (B_d) $ can provide
interesting bounds on BSM models:
\begin{eqnarray}
   \frac{\tau (B_s) }{\tau (B_d) } & = & 1 
   + \left[ \Gamma (B_d)- \Gamma (B_s)\right]^{\rm HQE}  \tau (B_s)^{\rm Exp} 
   + \left[ \Gamma (B_d)- \Gamma (B_s)\right]^{\rm BSM}  \tau (B_s)^{\rm Exp} \, .
\end{eqnarray}
BSM contributions could hereby stem from enhanced  $b \to s \tau^+ \tau^-$ transitions, from new decay channels in 
the baryogenesis models  discussed earlier \cite{Alonso-Alvarez:2021qfd} or from new effects in non-leptonic
tree level decays, see e.g. the talk of Matthew Kirk for the case of  $b \to c \bar{c} s $ transitions.
Bounds on $b \to s \tau^+ \tau^-$ obtained in that way seem to be competitive with current direct experimental 
bounds.}
\item[$\diamond$] Despite the promising opportunities of a precise measurement of the ratio  
$\tau (B_s)  / \tau (B_d) $, I have to point out a possible (temporary) showstopper. 
Recently the contribution of the Darwin term\footnote{Charles Galton Darwin, a grandson of THE Charles Darwin,
identified a relativistic correction to the Hydrogen atom, whose generalisation is active in the HQE.} 
was for the first time determined for non-leptonic decays 
 \cite{Lenz:2020oce,Mannel:2020fts,Moreno:2020rmk} and the Wilson coefficient $C_D$ turned out to be large and
 negative. The structure of the contribution of this term to the lifetime ratio looks like
 \begin{eqnarray}
   \frac{\tau (B_s) }{\tau (B_d) } & = & 1 + ...
   + \frac{C_D}{m_b^3} \left[ \rho_D^3 (B_d)-  \rho_D^3 (B_s)\right]  \tau (B_s)^{\rm Exp} + ...
\, .
\end{eqnarray}
The numerical value $ \rho_D^3 (B_d)$ is known from fits of moments of the inclusive semileptonic decays of 
$B_d$ or $B^+$ mesons, see e.g.  \cite{Alberti:2014yda,Bordone:2021oof}. The value of  $ \rho_D^3 (B_s)$ is unknown, naive
estimates using equations of motions and neglecting higher order terms in an $1/m_b$ expansion, result in
a huge amount of  $SU(3)_F$ breaking of  
about $50\%$ $SU(3)_F$, which would sizably change the theory prediction of the lifetime ratio towards lower values of $\Gamma_s$. Here clearly more work is necessary.
\end{itemize}

\section{Surprises/Underdogs}
\label{sec:surprise}

\subsection{Very rare to impossible decays}
A measurement of lepton flavour violating decays like $B_{(s)} \to \mu \tau$,  $B_{(s)} \to K \mu \tau$,  and also equivalent $D$ meson decays (see the  talks of Bhargav Joshi, Elena Occo, Alexei Sibidanov)
will be a clear signal of BSM physics. Here a further increase in the  experimental precision has of course an interesting  discovery potential.
\\
There is also a number of lepton  decays that fall in the same category of {\it very rare to impossible decays}, as e.g. $ \tau \ \to \mu \mu \mu  $ (see the  talks of Bhargav Joshi,  Emilie Passemar, A. Cerri, Paul  Feichtinger) - again a discovery of a sizeable value of the branching ratio would be an unambiguous signal for BSM effects.
\\
 {\color{blue}
 A future precise measurement of
$B_c \to \tau \nu $  will enable a direct determination of $V_{cb}$ depending  only on the decay constant $f_{B_c}$, providing thus an interesting cross-check of the $V_{cb}$ puzzle discussed in the beginning.
\\
Finally to further test the Cabibbo anomaly, precise measurements of $V_{us}$ from $\tau$ decays would not require
form factors as non-perturbative input and thus the CKM element extraction will be less theory dependent. 
}
\subsection{Inclusive semileptonic decays}
\label{sec:incl}
\begin{itemize}
\item[$\diamond$]  {\color{blue}
Fit of the moments of semileptonic inclusive $B_s$ decays, e.g. at Belle II at an $\Upsilon(5S)$ run or at LHCb(?) in order to extract a value of the matrix element of the Darwin operator for the $B_s$ meson, i.e.
$ \rho_D^3 (B_s)$. This would dramatically increase the theory precision of the lifetime ratio $\tau (B_s)  / \tau (B_d) $ and further enable the interesting physics program discussed in Section \ref{sec:tauBs}, see also the talks of Lu Cao, Paolo Gambino and Mortiz Bauer.}
\item[$\diamond$]  {\color{blue}
Fit of the moments of semileptonic inclusive $D^0$, $D^+$ and $D_s^+$,  decays, e.g. at 
BESIII or Belle II  in order to extract a value of the matrix element of the kinetic operator, the chromomagnetic operator and the Darwin operator for these $D$ mesons -  this would considerably increase the theory precision
of lifetime predictions of $D$ mesons, see below.

Lifetimes of charmed mesons are known precisely and they have
recently been confirmed by Belle II, \cite{Belle-II:2021cxx}, see the talk by Giulia Casarosa. In contrast to the $B$ system, the lifetimes of the $D$ mesons can differ significantly among each other and we find \cite{ParticleDataGroup:2020ssz}
\begin{equation}
    \frac{\tau (D^+)}{\tau (D^0)} = 2.54 \pm 0.02
    \, , 
    \hspace{1cm}
    \frac{\tau (D_s^+)}{\tau (D^0)} = 1.20 \pm 0.01
    \, . 
    \end{equation}
From a theoretical point of view these decays are described by the HQE, according to which we can express the total decay rate as a Taylor series
in inverse powers of the charm quark mass
\begin{equation}
\Gamma(D) = 
\Gamma_3  +
\Gamma_5 \frac{\langle {\cal O}_5 \rangle}{m_c^2} + 
\Gamma_6 \frac{\langle {\cal O}_6 \rangle}{m_c^3} + ...  
 + 16 \pi^2 
\left( 
  \tilde{\Gamma}_6 \frac{\langle \tilde{\mathcal{O}}_6 \rangle}{m_c^3} 
+ \tilde{\Gamma}_7 \frac{\langle \tilde{\mathcal{O}}_7 \rangle}{m_c^4} + ... 
\right) .
\label{eq:charm}
\end{equation}
The coefficients $\Gamma_i$ denote perturbative Wilson coefficients and the $\langle {\cal O}_i \rangle$ are 
non-perturbative matrix elements of operators of mass dimension $i$.
In contrast to the quantities discussed in 
Section \ref{sec:CPVcharm} (exclusive hadronic decays or $D$ mixing with extreme GIM cancellations)  inclusive decay rates (including lifetimes) are much simpler quantities and therefore
 they are better suited for testing convergence properties of our theory tools in the charm sector.

In \cite{King:2021xqp} we performed a comprehensive study of 
total decay rates, lifetime ratios and semileptonic branching ratios (see  \cite{BESIII:2021duu} and the talk of Alex Gilman for interesting new results from BESIII for inclusive semileptonic $D_s^+$ decays) of $D$ mesons and we find in general an agreement of the HQE with data, albeit with huge uncertainties, see Fig. \ref{fig:charm}.

 \begin{figure}
\centering
\includegraphics[width=0.9\textwidth]{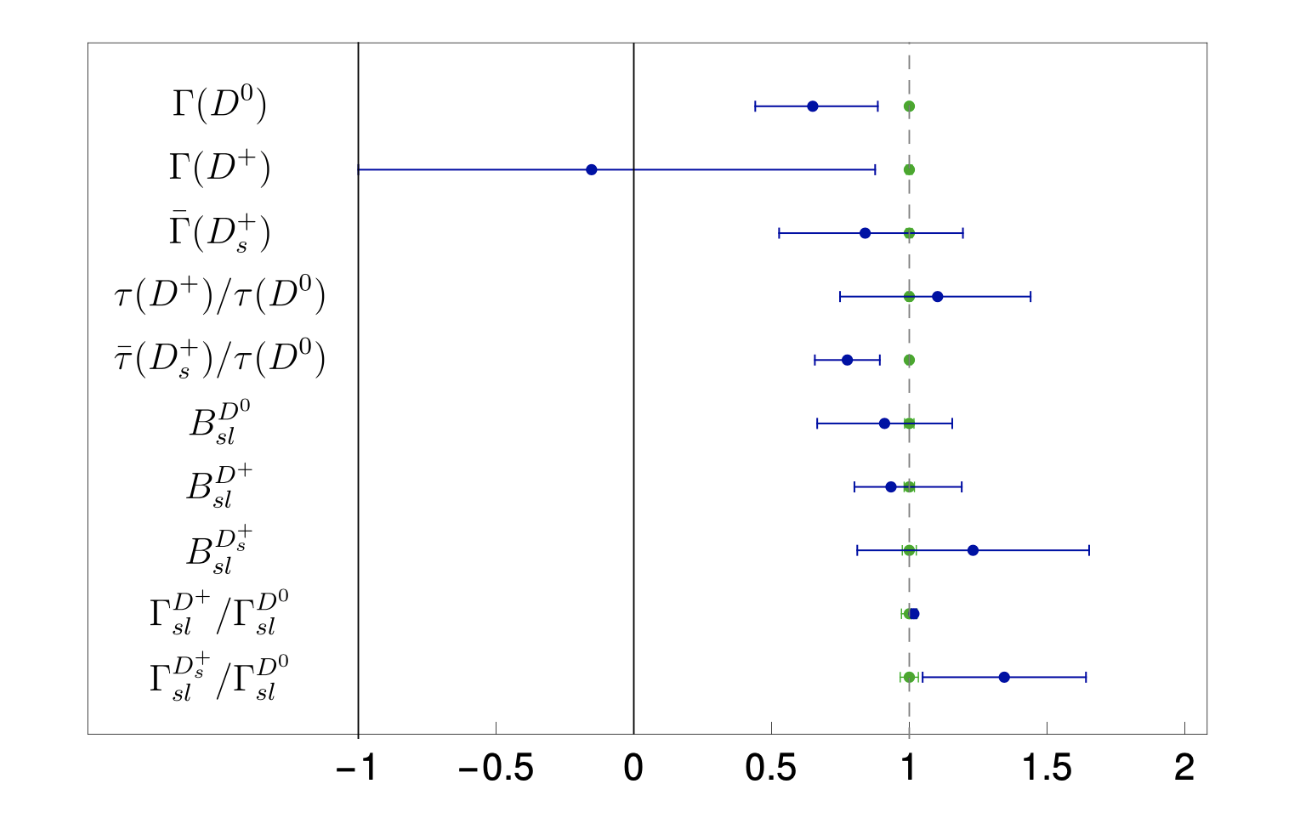}
\caption{\label{fig:charm} 
A comparison of the HQE prediction for the charm observables in the kinetic scheme  (blue)  with the corresponding experimental data (green).
    All the quantities are normalised to the corresponding experimental central values.
}
\end{figure}

In order to identify the dependence of the $D$ meson decay
rates on the different input parameters, we present the
formulae in the following form \cite{King:2021xqp}
\begin{eqnarray}
\Gamma (D_s^+) 
& = & 
6.15 \, \Gamma_0
\biggl[1 + 0.48 
- 0.15 \, \frac{\mu_{\pi}^2 (D_s)}{0.555 \, \rm GeV^2}
+ 0.01 \, \frac{\mu_{G}^2(D_s)}{0.36 \, \rm GeV^2}  
+ 0.42 \, \frac{\rho_{D}^3(D_s)}{0.110 \,  \rm GeV^3}
\nonumber \\[2mm] 
& &
\quad 
- \!\!\! \underbrace{0.20}_{\rm dim-6, VIA} \!\!\!
- \, 0.161 \,  \frac{\delta {\tilde B}_1^s }{0.02}
+ 0.157 \,  \frac{{\tilde B}_2^s}{0.02}
+ 0.089 \,  \frac{\tilde \epsilon_1^s}{-0.04}
- 0.122 \,  \frac{\tilde \epsilon_2^s}{-0.04}
+ \!\!\! \underbrace{0.10}_{\rm dim-7, VIA}
\nonumber \\[2mm]
& &
\quad 
- \, 0.0064 \,  r^{qs}_{1}
- 0.0007 \,  r^{qs}_{2} 
- 0.0036 \,  r^{qs}_{3}
+ 0.0012 \,  r^{qs}_{4}
\biggr] \, .
\label{eq:Gamma-Ds}
\end{eqnarray}
For our purposes here, we are only interested in the first line,
while the parameter in the second and third line can be determined via HQET sum rules estimates as done in \cite{Kirk:2017juj}.
The non-perturbative parameter in the first line $\mu_{\pi}^2 (D_s)$,
$\mu_{G}^2(D_s)$
and
$\rho_{D}^3(D_s)$ are essentially unknown. Assuming heavy quark symmetry we can use the corresponding values from the $B_d$ and $B^+$ system, which were determined by a fit of the moments of the semileptonic $B$ Decays, see e.g. \cite{Alberti:2014yda,Bordone:2021oof}. Using the $B$ system
values, we find that the effective coefficients of the non-perturbative parameter are quite large - e.g. the Darwin term is expected to give a correction to the free charm quark decay of about $40 \%$.

Here an experimental extraction of $\mu_{\pi}^2 (D_s)$,
$\mu_{G}^2(D_s)$
and
$\rho_{D}^3(D_s)$ 
from inclusive semileptonic $D_s^+$ meson decays would considerably improve the precision of the HQE predictions for the $D$ meson - this might also shed some light into the theory tools used for CP violation in $D$ decays and for $D$ mixing.
}
 \end{itemize}
 \subsection{Exclusive non-leptonic tree-level decays: }
\label{sec:NLTD}
{\color{blue}
We have already mentioned several times BSM effects
in non-leptonic tree-level decays. 
Now I would like to focus on some recent developments
related to colour-allowed decays starting with
the work in \cite{Bordone:2020gao}\footnote{See also
the excellent talks in this topical workshop:
https://indico.scc.kit.edu/event/2352/}, where
previous observations \cite{Huber:2016xod} 
of deviations of  experimental data \cite{LHCb:2021qbv} for branching
ratios from theory predictions based on QCD
factorisation \cite{Beneke:2000ry}
were confirmed.
In particular the new study finds for the decay 
  $\bar{B}_s \to D_s^+  \pi^-$  a discrepancy of
 about  four standard deviations. 
 This decay is triggered by the CKM leading
 tree-level decay $b \to c \bar{u} d$ and it is a
 colour-allowed decay, which cannot have corrections
 due to penguins or annihilation contributions - thus
 a perfect playground for QCD factorisation. 
 The SM results have been confirmed in
 \cite{Cai:2021mlt,Endo:2021ifc}, 
 QED corrections to these decays were studied in
 \cite{Beneke:2021jhp}
 and BSM explanations were 
 considered in \cite{Cai:2021mlt,Iguro:2020ndk}
 and challenged by collider bounds in \cite{Bordone:2021cca}.
 Here further more precise experimental determinations of the branching ratios of decays like 
  $\bar{B}_s \to D_s^{(*)+}  \pi^-$ and  $\bar{B}_s \to D^{(*)+}   K^-$ would be beneficial for clarifying this discrepancy.
 New effects in theses decays will also modify the quantities $\tau (B_q)$,  $\Delta \Gamma_q $ and  $a_{sl}^q$, that have been discussed earlier.
 \\
 It is interesting to note that the decay  $\bar{B}_s \to D_s^+  \pi^-$  was also used for the most precise measurement of the mass difference of neutral $B_s$, $\Delta M_s$ by the LHCb Collaboration \cite{LHCb:2021moh},
 which agrees well with the much less precise SM prediction \cite{DiLuzio:2019jyq}. This agreement leads to strong bounds on BSM models trying to explain the flavour anomalies, see e.g.  \cite{DiLuzio:2019jyq,DiLuzio:2017fdq}.
 \\
 Maybe even more interesting: since within the SM  $\bar{B}_s \to D_s^+  \pi^-$  is a flavour specific-decay, as
 defined in Section 2.3, this channel might also be used in experiment for determining the flavour-specific
 CP asymmetries. Beyond the SM the decay  $\bar{B}_s \to D_s^+  \pi^-$ might also have some new CP violating contributions, leading to direct CP violation in this decay and thus not being flavour-specific anymore.
 Therefore a measurement of the asymmetry  defined in Eq.(\ref{ea:def_asl}) with the final state  $ D_s^+  \pi^-$might yield
 results that differ significantly from a future more precise measurement of $a_{sl}^s$ - this would be an unambiguous signal for new physics.}
  \\
 \\
 In that respect one has probably also to mention the 
 longer standing $B \to K \pi$ puzzle, where in contrast to the decays discussed above, we have here significant contributions of penguins, which are responsible for the existence of CP asymmetries. Some related  experimental results were discussed in the talk of 
 Stefano Perazzini. Despite some significant theory improvements \cite{Bell:2020qus} the puzzle seems to persist - despite an updated comprehensive phenomenological analysis within the framework of QCD factorisation is still missing.

\section{Conclusion}
I hope  my experimental friends find anything useful in these proceedings and I was trying hard to come up with additional, beyond-theory motivation for working on the above listed observables. Here it is:
in case of having a 
measurement of $a_{sl}^s$ with an uncertainty of $70 \cdot 10^{-5}$, 
 a measurement of inclusive semileptonic $B_s$
 meson decays leading to a precision of 
 $ \rho_D^3 (B_s)$  of $25 \%$,
 a measurement of inclusive semileptonic $D$
 meson decays leading to a precision of 
 $ \rho_D^3 (D_s^+)$ with a relative precision
 of $40 \%$ 
 and  a measurement of  $\tau (B_s)  / \tau (B_d) $ at the two per 
 mille level accuracy, 
  I promise to organise an opportunity for presenting these results at the most inspiring venue we were allowed to use so far for scientifc contferences.\footnote{https://www-archive.ph.ed.ac.uk/heavyflavour2016/
  }

\section*{Acknowledgment}
Thanks to  Alexander Khodjamirian, Rusa Mandal, Thomas Mannel, Aleksey Rusov and Gilberto Tetlalmatzi-Xolocotzi for suggesting observables
and
to Tim Gershon and Matthew Kirk for helpful discussions and
to Maria Laura Piscopo and Aleksey Rusov for proof-reading.

\bibliographystyle{ieeetr}
\bibliography{References}

\end{document}